\documentclass[useAMS]{mn2e}
\usepackage{epsfig}

\newcommand{\kms}{km\,s$^{-1}$}

\newcommand{\HeI}  {He\,{\sc i}}

\newcommand{\KI}  {K\,{\sc i}} 
\newcommand{\NaI}  {Na\,{\sc i}} 
\newcommand{\OI}   {O\,{\sc i}}

\newcommand{\Halpha} {H$\alpha$}

\begin{document}

\title[The donor star in IY\,UMa]
  {The detection of the donor star in IY\,UMa}
\author[D. J. Rolfe et al.]
  {Daniel J. Rolfe$^{1,2}$, Timothy M. C. Abbott$^{3,4}$ and Carole A. Haswell$^2$\\
$^1$Department of Physics \& Astronomy, University of Leicester, University Road, Leicester, LE1 7RH, UK.\\
$^2$Department of Physics \& Astronomy, The Open University, Walton Hall, Milton Keynes, MK7 6AA, UK.\\
$^3$Cerro Tololo Inter-American Observatory, Casilla 603, La Serena, Chile\\
$^4$Nordic Optical Telescope, Roque del Los Muchachos \& Santa Cruz de La Palma, Canary Islands, Spain}
\date{Accepted. Received}
\pagerange{\pageref{firstpage}--\pageref{lastpage}}
\pubyear{2002}

\maketitle \label{firstpage}
\begin{abstract}
  We present the results of a search for the donor star in the high
  inclination SU\,UMa type cataclysmic variable IY\,UMa. We detect
  absorption features in the near infrared consistent with an M type
  dwarf donor star. Using the skew mapping technique to exploit the
  velocity information provided by the 8183--8194\AA\, Na\,I
  absorption doublet, we locate the absorption at the expected donor
  velocity of IY\,UMa.
\end{abstract}

\begin{keywords}
  stars: novae, cataclysmic variables, stars: individual: IY\,UMa,
  stars: dwarf novae, techniques: spectroscopic
\end{keywords}

\section{Introduction}

IY\,UMa is a dwarf nova cataclysmic variable (CV). Patterson et al.
(2000) (hereafter P2000) and Rolfe, Haswell and Patterson (2001)
present detailed photometric studies of the January 2000 superoutburst
and superhump phenomenon. Spectroscopic observations have been
published, revealing a bright hotspot and deep white dwarf absorption
lines during quiescence (Rolfe, Abbott \& Haswell 2001, Patterson et
al.  2000, Rolfe et al.  2002); an eccentric disc during superoutburst
(Wu et al. 2001); and lines powered by reprocessed boundary layer
emission during outburst (Rolfe, Abbott \& Haswell 2002 and Rolfe et
al. 2002).

The orbital period of IY\,UMa is 1.77 hours, and Patterson et al.
(2000) estimated $M_{wd}=0.86\pm0.11 M_\odot$, $M_{donor}=0.12\pm0.02
M_\odot$ and orbital inclination $i=86\fdg8 \pm 1\fdg5$. Direct
measurements of orbital parameters are desirable. A model-independent
solution can be obtained if the orbital period and white dwarf eclipse
width (already accurately measured for IY\,UMa) can be combined with
measurements of the donor and white dwarf orbital velocities. The
orbital period suggests a spectral type of M4.2 to M6.3 for the donor
in IY\,UMa (Equation 4 in Smith \& Dhillon (1998)). Thus absorption
features in the far optical and near infrared are expected cf. Z~Cha
and HT~Cas (Wade \& Horne 1988 and Marsh 1990). We obtained
spectroscopic observations of IY\,UMa to look for these absorption
features in the wavelength range 7000--8300\AA.

\section{Observations and data reduction} 

Observations were taken on 2001 Jan 3 and 4 using ALFOSC on the Nordic
Optical Telescope. A few calibration images were taken on Jan 7. A
slit width of 1.2$''$ was used with GRISM 8 yielding a FWHM resolution
of $\sim$6.2\AA\, and dispersion 1.24\AA~per pixel. We discuss only
the red end of the spectra here (7000--8350\AA). 39 360-second
exposures of IY\,UMa and a nearby comparison were obtained, along with
exposures of several M type dwarfs for use as template stars (listed
in Table \ref{IYUMa:Spectroscopy:TemplateTable}). Frequent arc lamp
exposures provided wavelength calibration (giving 9\,\kms\, r.m.s.
velocity variation in the sky lines) while HD\,93521 served as a flux
standard and telluric standard.

\begin{table}
\caption{M dwarf template stars}
\label{IYUMa:Spectroscopy:TemplateTable}
\begin{center}
\begin{tabular}{lccl}
%\hline
Object & Spectral type & Reference\\
\hline
GJ~1154, Zkh~176    &  M5V            &  Zakhozhaj (1979)\\
GJ~228B, Zkh~84     &  M4V            &  Zakhozhaj (1979)\\
BD+28~2110          &  M3V            &  Upgren (1962)\\
CTI~092053.7+280101 & M1.5V           &  Kirkpatrick et al. (1994)\\
%\hline
\end{tabular}
\end{center}
\end{table}

Data reduction was carried out using {\sc iraf}. The CCD suffers from
fringing but normalized lamp flats were used to flat-field the images
and removed all detectable traces of the fringing pattern. The {\sc
  telluric} package in {\sc iraf} was used to correct for telluric
features. Apart from the strongest feature around 7600\AA, the
telluric correction worked quite well.

Conditions were not photometric so flux calibration of these red
spectra was not possible. However, the skew mapping technique relies
on normalized spectra. Each spectrum was normalized by fitting a 4th
order polynomial to the continuum regions and dividing by this.

\section{Average spectra}

\begin{figure*}
  \psfig{file=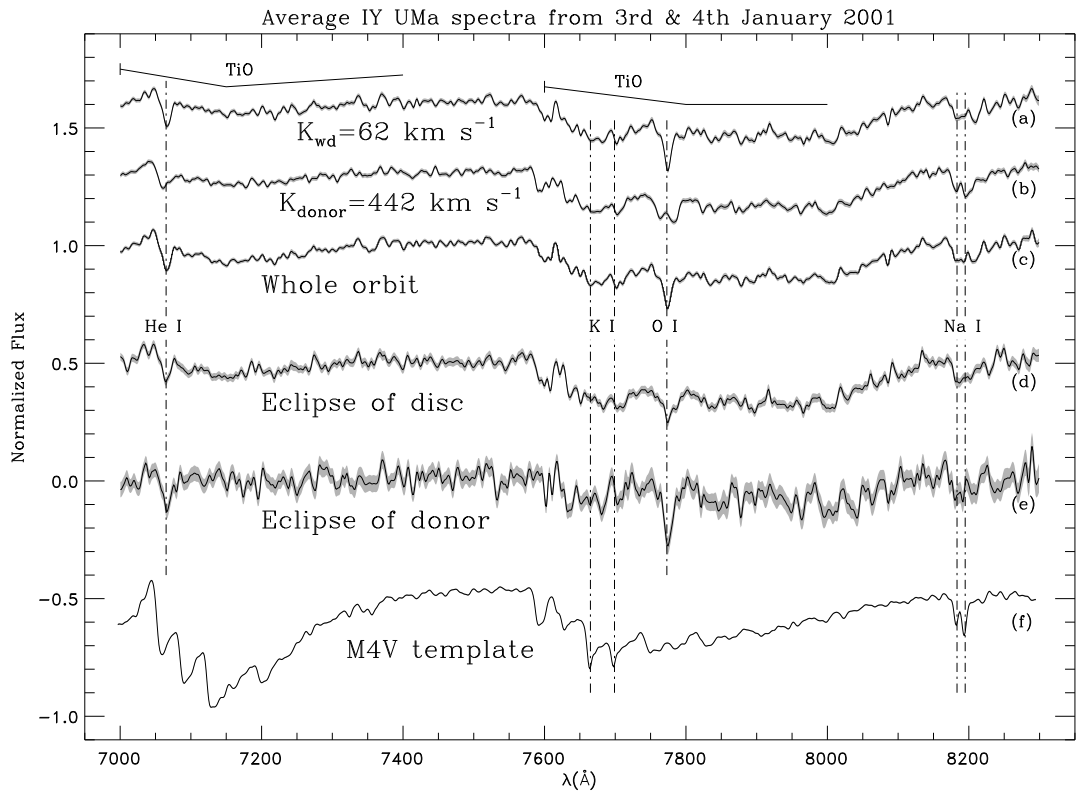,height=0.8\textwidth,angle=90}
  \caption{
    Average red spectra of IY\,UMa on 3rd \& 4th January 2001. (a)
    shows average in rest frame of white dwarf, (b) in rest frame of
    donor, (c) in frame of observer. (d) shows average when donor
    eclipses disc and white dwarf, (e) when disc obscures donor. (f)
    shows M4V template spectrum.  Shaded regions show error bars.
    }
  \label{IYUMa:Spectroscopy:RedSpectraFig}
\end{figure*}

Various different averages of the IY\,UMa spectra are shown in Figure
\ref{IYUMa:Spectroscopy:RedSpectraFig}a--e along with the M4V dwarf
spectrum in Figure \ref{IYUMa:Spectroscopy:RedSpectraFig}f. Each
plotted spectrum has been Gaussian smoothed with FWHM of 4 pixels
(less than the FWHM spectral resolution), and shifted in flux by a
multiple of 0.5 units.
  
Figure \ref{IYUMa:Spectroscopy:RedSpectraFig}c shows the average of
all 39 IY\,UMa spectra. The most obvious features are the \HeI~emission
and core absorption at 7065\AA, the 7773\AA~\OI~absorption, the broad
absorption band around 7600\AA~to 8200\AA~and the absorption line
around 8200\AA. The \HeI\, is from the disc.

The broad 7600\AA--8200\AA~feature matches the TiO absorption band
seen in the M dwarf spectra, where it is strongest in the M5V dwarf.
Figures \ref{IYUMa:Spectroscopy:RedSpectraFig}d \&
\ref{IYUMa:Spectroscopy:RedSpectraFig}e show average spectra of
IY\,UMa during eclipse (orbital phases -0.1--0.1) and where the donor
should be partially obscured by the disc (phase 0.42--0.58 assuming a
quiescent disc radius of 0.25$a$ (Stanishev et al. 2001)). The TiO
absorption is clearly much stronger (by a factor $\sim$2) when the
donor eclipses the disc than when the disc obscures the donor,
confirming the identification of this feature as the TiO band
absorption in the donor.

The 7773\AA\, \OI\, triplet has been seen in many CVs. A study of 65
CVs by Friend et al. (1988) reveals this feature in absorption in
several systems during outburst, but in emission during quiescence.
HT~Cas also shows this feature in emission during quiescence
\cite{Marsh:1990}. Friend et al. (1988) attribute this to line
emission from an optically thin disc in quiescence, and absorption
from an optically thick disc in outburst. Two high inclination
systems, Z~Cha \cite{WadeEt:1988} and V2051~Oph \cite{FriendEt:1988},
show a broad emission from the \OI\, triplet, superimposed with a
narrower absorption component. It is possible that a broad weak
emission component is also present in these IY\,UMa spectra. Friend et
al. (1988) suggest that the narrow absorption feature is absorption
of white dwarf emission by the disc, something which we might expect
in high inclination systems. That this feature arises from the disc
and certainly not from the donor is confirmed by Figures
\ref{IYUMa:Spectroscopy:RedSpectraFig}a \& b, where each individual
spectrum was Doppler-shifted into the rest frame of the white dwarf or
donor before adding together the spectra. The velocities of the donor
and white dwarf were assumed to be $K_{2}=442$~\kms and
$K_{1}=62$~\kms\, respectively, calculated using the P2000 orbital
parameters. In the rest frame of the white dwarf the \OI~feature
remains sharp, while in the donor star frame it splits into two
absorption dips. In addition, the depth of this absorption is
significantly stronger when the disc is in front of the donor (Figure
\ref{IYUMa:Spectroscopy:RedSpectraFig}e) than when the donor eclipses
the disc (Figure \ref{IYUMa:Spectroscopy:RedSpectraFig}d).

The 8200\AA~absorption feature closely matches the 8183--8194\AA\,
\NaI\, doublet seen in all four dwarf template spectra. In the average
spectrum (Figure \ref{IYUMa:Spectroscopy:RedSpectraFig}c) it does not
have the same doublet structure as seen in the template spectra, but
when shifted to the donor rest frame (Figure
\ref{IYUMa:Spectroscopy:RedSpectraFig}b) the doublet structure becomes
apparent, clearly matching the laboratory wavelengths. This feature is
also more significant (considering the flux errors) during disc
eclipse (Figure \ref{IYUMa:Spectroscopy:RedSpectraFig}d) than when the
disc passes in front of the donor (Figure
\ref{IYUMa:Spectroscopy:RedSpectraFig}e). This is a clear detection of
the \NaI~doublet, a feature which has been exploited to measure $K_2$
in other CVs, c.f. Wade \& Horne (1988), Friend et al. (1988, 1990a,
1990b) and Marsh (1990).

The TiO absorption band around 7000\AA--7400\AA\, is also more
prominent when the donor obscures the disc than when the disc obscures
the donor. The \KI\, lines at 7665\AA\, and 7699\AA\, are not clearly
seen, but there are weak features close to the noise level which could
be tentatively identified as these lines.

Weaker \NaI\, and TiO absorption around phase 0.5 than at other phases
has been seen before in dwarf novae, e.g. Z~Cha \cite{WadeEt:1988},
HT~Cas \cite{Marsh:1990} and SS~Cyg \cite{HessmanEt:1984}, and was
attributed to a non-uniform distribution of absorption across the
surface of the donor. The suggested cause of this non-uniform
distribution is the filling-in of absorption features on the inner
face of the donor by irradiation from the disc and/or white dwarf.
This, rather than eclipse of the donor by the disc, might also be the
cause of the weakened features around phase 0.5 in IY\,UMa.

The absorption feature at 7600\AA\, is probably a combination of
residual atmospheric absorption and the 7600\AA\, feature seen in M
dwarf spectra. There is some evidence of over-correction of the
telluric absorption in the M5V template around 8230\AA.

We see no evidence of Paschen emission around 8200\AA\, seen in some
CVs e.g. TT~Ari and V603~Aql \cite{FriendEt:1988}.

The donor star in IY\,UMa has been detected, showing all the expected
features of a late M dwarf star. Without flux calibrated spectra, the
relative strengths of the absorption features cannot be used to
estimate more precisely the spectral type of the donor. The low
signal-to-noise ratio prevents us from exploiting the \NaI\, doublet
to measure $K_2$ directly from the spectra, so we adopt the skew
mapping technique \cite{SmithEt:1993} to locate the donor in velocity
space.

\section{Skew mapping}

Skew mapping is the best method for detecting the donor and
determining its velocity when the absorption features are weak. This
technique has been successfully employed for several systems (e.g.
Smith, Cameron and Tucknott (1993), Smith, Dhillon and Marsh (1998)
and recently in VY Aqr (Littlefair et al 2000)). Skew mapping co-adds
spectra to increase signal-to-noise, while also correcting each
spectrum for the orbital motion of the donor, to avoid smearing out
donor absorption features. Skew mapping extends the technique used in
producing Figure \ref{IYUMa:Spectroscopy:RedSpectraFig}b (where $K_2$
was assumed), by cross-correlating each such co-added spectrum with a
template donor star spectrum, and finding the combination of template
star and donor velocity at which the cross-correlation peak is
strongest. This process is equivalent to producing a trailed
``cross-correlation spectrum'', and then back-projecting this trail
into velocity space, producing a cross-correlation map in
velocity-space (the skew map) which should show a peak at the velocity
of the donor star.

The normalized spectra of the template M dwarfs and IY\,UMa were
continuum subtracted, and were Gaussian smoothed with FWHM of 2.5
pixels to increase S/N without losing any spectral information. Each
IY\,UMa spectrum, $F(\lambda)$, was cross correlated with the template
spectrum, ${T}(\lambda)$, for a range of velocity shifts, $v$, using
correlation coefficient $C_v$ defined as
\begin{equation}
\label{IYUMa:Spectroscopy:SkewMappingCoeff}
C_v=\frac
{\sum_{\lambda=\lambda_{min}}^{\lambda_{max}}F(\lambda){T'}_v(\lambda)}
{\left(\sum_{\lambda=\lambda_{min}}^{\lambda_{max}}F(\lambda)^2 \sum_{\lambda=\lambda_{min}}^{\lambda_{max}}{T'}_v(\lambda)^2\right)^\frac{1}{2}},
\end{equation}
where ${T'}_v(\lambda)=T(\frac{\lambda}{1+v/c})$. This definition of
${T'}_v(\lambda)$ simply has the effect of red-shifting the template
spectrum by velocity $v$ before cross-correlating. The denominator of
$C_v$ is chosen so that if the template and donor spectra are
identical, the value of $C_v$ at the peak will be 1, independent of
any simple scaling of the flux of either spectrum. The correlation
trails thus produced were then transformed to velocity-space using the
Fourier-filtered back-projection technique as implemented in Tom
Marsh's {\sc molly} software.  The strength of the skew map at a given
velocity therefore shows how well the object spectra match a star of
the template's spectral type orbiting at that velocity.

Only the \NaI~doublet wavelength range was used in the
cross-correlation, since this is the only sharp donor absorption
feature detected, and will therefore provide all the velocity
information. Skew maps were also produced using the spectral range
$\lambda=$ 7045--8240\AA\, (with the \OI~and \HeI~lines masked out)
which uses all the detected donor star features, but this simply
smears out the donor star in the skew map, probably a result of the
broad TiO absorptions showing a strong correlation but low velocity
sensitivity.  The normalization of the spectrum around the
\NaI~doublet was improved by fitting a straight line to the continuum
either side of the doublet. The skew map also depends upon the
relative systemic velocity of IY\,UMa and the template star. Neither
of these is yet known, although measurements of \Halpha\, emission
lead to estimates for IY\,UMa of $\gamma=$ 13.6$\pm$1\,\kms (Rolfe et
al. 2002) and $\gamma=$ -4$\pm$32\,\kms (Wu et al. 2001). Such
estimates should be treated with caution, however, since the
determination of systemic velocity from emission lines is notoriously
unreliable: effects such as phase-dependent occultation/absorption of
emission and temporal variations in emission can produce incorrect
results. By cross-correlating the M3V template whose systemic velocity
is 1\,\kms\, \cite{Evans:1967} with the other four templates, the
systemic velocities of these stars were determined. The skew maps for
each template were produced for a range of values of the IY\,UMa
systemic velocity, $\gamma_{\rm IYUMa}$, between -100 and 300~\kms
(the low resolution of our spectra limits the accuracy of our systemic
velocity determinations).  This was identified as a suitable range for
$\gamma_{\rm IYUMa}$ after initially studying the range -750--750\kms.
This procedure enables a determination of the value which produces the
clearest peak. We do not consider this a determination of the systemic
velocity of IY\,UMa.

\begin{figure}
  \centerline{\psfig{file=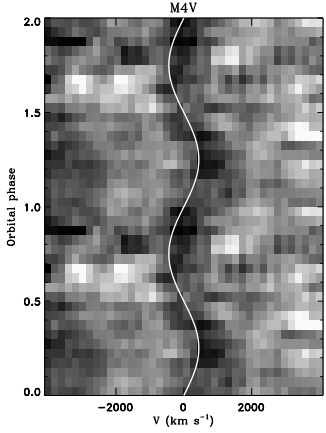,width=0.5\columnwidth}}
  \caption{Cross-correlation trail for the M4V template}
  \label{IYUMa:Spectroscopy:SkewmapTrails}
\end{figure}

The orbital phase range 0.42 to 0.58 was not included when producing
the back projections to avoid the weakened/obscured absorption lines
around this phase interfering with the results. The results from using
the entire phase range are considered at the end of this section.

Figure \ref{IYUMa:Spectroscopy:SkewmapTrails} shows a phase-folded and
-binned and velocity-binned correlation trail for the M4V template.
%The trail was folded and binned into 20 orbital phase bins, and into
%100 velocity bins in the range -4000~\kms\, to 4000~\kms.
All four templates yield almost identical trails; at this low
resolution and signal-to-noise, differences in structure in the
\NaI~doublet between the spectral types are not going to be
distinguishable in the IY\,UMa observations. The trail shows an
S-wave, corresponding to a localized region in velocity space at which
the IY\,UMa spectrum best matches the sodium feature. The peak-peak
amplitude of this feature is about 1000\,\kms, and is at minimum
excursion around orbital phase 0. This phasing and amplitude is in
agreement with the predicted orbital motion of the donor (plotted in
white in Figure \ref{IYUMa:Spectroscopy:SkewmapTrails}) using the
P2000 orbital parameters (line-of-sight donor velocity semi-amplitude
442~\kms). The velocities are heliocentric and corrected for the
template systemic velocity.

\begin{figure}
\centerline{\psfig{file=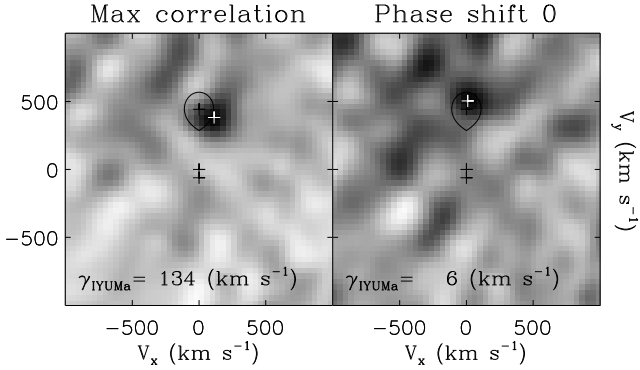,width=\columnwidth}}
\caption{
  IY\,UMa skew maps using the M4V template. The teardrop shows the
  expected velocity of the donor, and the black crosses show the white
  dwarf, system and donor centre of mass velocities. The white cross
  is the measured peak maximum.}
\label{IYUMa:Spectroscopy:SkewmapsFig}
\end{figure}

The trails were transformed to velocity space. The resulting skew maps
have a clear maximum close to the predicted location of the donor star
using the P2000 parameters. The exact location of the maximum depends
on the assumed IY\,UMa systemic velocity, $\gamma_{\rm IYUMa}$. Two
skew maps for the M4V template are shown in Figure
\ref{IYUMa:Spectroscopy:SkewmapsFig}.  The map in the left panel uses
the value of $\gamma_{\rm IYUMa}$ which produces the strongest peak
(i.e.  max correlation), while that on the right corresponds to the
$\gamma_{\rm IYUMa}$ which produces a peak with no x-component of
velocity (as expected for the donor). These values of $\gamma_{\rm
  IYUMa}$ were determined as described below.

\begin{figure}
\centerline{\psfig{file=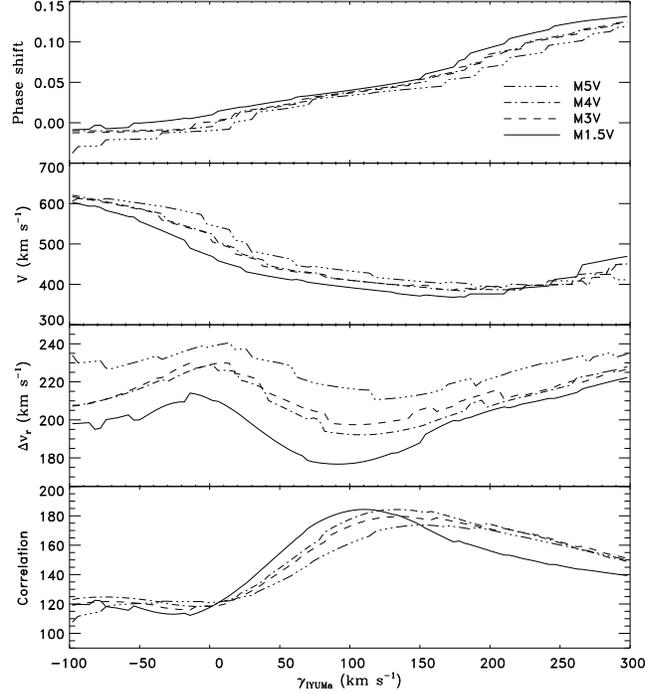,width=\columnwidth}}
\caption{
  Comparison of skew maps for each template.}
\label{IYUMa:Spectroscopy:CompareSkewmaps}
\end{figure}

The velocity and width of the peak was determined for each image by
fitting a 2-dimensional Gaussian of the form
$$C_{\vec{v}}=C_0+C_1\exp\left( -\frac{(v_x-v_{x,max})^2}{2\Delta
    v_x^2}-\frac{(v_y-v_{y,max})^2}{2\Delta v_y^2} \right)$$
where
$C_{\vec{v}}$ is the value of the skew map at velocity $(v_x,v_y)$.
Figure \ref{IYUMa:Spectroscopy:CompareSkewmaps} shows the variation in
position and shape of the skew map peak as a function of $\gamma_{\rm
  IYUMa}$.  The phase shift is the angular offset of the peak from
$v_x$=0 measured about the centre of mass, i.e.
$\frac{1}{2\pi}\arctan\frac{v_{x,max}}{v_{y,max}}$. The velocity
amplitude is $V=\sqrt{v_{x,max}^2+v_{y,max}^2}$. The peak width is
measured using $\Delta v_r=\sqrt{\Delta v_x^2+\Delta v_y^2}$. The
correlation of the peak is taken as $C_0+C_1$.

As $\gamma_{\rm IYUMa}$ increases from -100\,\kms~to 300\,\kms, the
peak moves from a velocity of about 600\,\kms and phase shift of
around -0.01 through the expected donor location to a lower velocity
around 400--450\,\kms and phase shift of around 0.13. This result is
true for for all four templates.  Only the peak size $\Delta v_r$ and
the peak correlation vary appreciably depending on the template used.
The difference in $\Delta v_r$ between templates should not be taken
as a measure of how tightly each template constrains the location of
the donor: it is more likely to result from differing seeing between
the different template observations. However, the variation of $\Delta
v_r$ with $\gamma_{\rm IYUMa}$ for each template is more useful,
telling us which value of the IY\,UMa systemic velocity gives us the
sharpest peak, and is therefore the most likely value for the systemic
velocity. This shows the peak to be narrowest for $\gamma_{\rm IYUMa}$
around 75 to 125\,\kms. The most important measure of which value of
$\gamma_{\rm IYUMa}$ is correct is that for which the peak correlation
is strongest.  This is around $\gamma_{\rm IYUMa}$=100 to 150\,\kms\,
for the four templates.  The peak correlation of the templates is
similar, which considering the low signal-noise and resolution of
these observations is not surprising: detailed differences in the
\NaI\, doublet between the spectral types have not been resolved.

The value of $\gamma_{\rm IYUMa}$ with the highest peak correlation
gives us one measure of the best fitting skew map. The corresponding
M4V skew map is shown in the left panel of Figure
\ref{IYUMa:Spectroscopy:SkewmapsFig}. The average phase shift for all
four of these maps is 0.044, with a spread in values of 0.001. The
360~second exposure time used corresponds to a phase resolution of
0.06, so this phase shift is consistent with the peak actually lying
along the line of centres of the two stars, where we expect the donor.
The accumulated error in the P2000 ephemeris at the time of these
observations is only 0.005 orbits. This then gives us a donor velocity
of $K_2$=383\,\kms\, with a spread of values of 6\,\kms.  If this
measured phase shift is real, then we should also consider values of
$\gamma_{\rm IYUMa}$ for which the phase shift is zero, placing the
donor along the line of centres. The corresponding skew maps (shown
for M4V in the right panel of Figure
\ref{IYUMa:Spectroscopy:SkewmapsFig}) give $K_2$=524\,\kms\, with a
spread of 19\,\kms. A possible cause of such a phase shift would be
irradiation of the donor by emission from the stream-disc impact,
filling-in the Na absorption mainly on the leading side of the donor,
thus moving the correlation peak in the direction seen.

As described earlier, these skew maps were produced omitting phase
range 0.42--0.58. As a check, the analysis was repeated using the
entire phase range. The skew maps corresponding to phase shift zero
show far less distinct peaks than in Figure
\ref{IYUMa:Spectroscopy:SkewmapsFig}, rendering this measurement less
reliable. The peak in correlation, the minimum value of $\Delta v_r$
and phase shift of zero occur for values of $\gamma_{\rm IYUMa}$ about
10--20\,\kms\, greater than before. The phase shift at maximum
correlation is now 0.057 with a spread of 0.003, as large as the phase
resolution of the data. The orbital velocity of the donor for maximum
correlation is now $K_2$=362\,\kms\, with a spread of 1\,\kms.

\section{Conclusions}
  
The donor star has been unambiguously detected in IY\,UMa. It is a
late type M dwarf, but the low signal-to-noise and resolution of these
observations prevents an accurate determination of the spectral type.
The donor velocity, $K_2$, lies in the range about 380--540\,\kms,
consistent with the value of 442\,\kms\, using the model-dependent
orbital parameters estimated from the photometric study in P2000.
Observations with a larger telescope, providing higher spectral and
temporal resolution and signal-to-noise would facilitate a direct and
accurate determination of both the spectral type and velocity of the
donor star in IY\,UMa.

\section{Acknowledgements}

The data presented here have been taken using ALFOSC, which is owned
by the Instituto de Astrofisica de Andalucia (IAA) and operated at the
Nordic Optical Telescope under agreement between IAA and the NBIfAFG
of the Astronomical Observatory of Copenhagen. The Nordic Optical
Telescope is operated on the island of La Palma jointly by Denmark,
Finland, Iceland, Norway, and Sweden, in the Spanish Observatorio del
Roque de los Muchachos of the Instituto de Astrofisica de Canarias.
DJR would like to thank Tom Marsh for the use of his {\sc molly}
software, and Ulrich Kolb and Brian Warner for useful comments. DJR
was supported by a PPARC studentship and the OU research committee.
CAH gratefully acknowledges support from the Leverhulme Trust
F/00-180/A.

\label{lastpage}

\end{document}